\pdfoutput=1
\documentclass[11pt]{article}

\usepackage{ACL2023}
\usepackage{times}
\usepackage{latexsym}
\usepackage{xcolor}
\usepackage{enumerate}
\usepackage{soul}
\usepackage{xcolor}
\usepackage{framed}
\usepackage{graphicx}
\usepackage{subfigure}
\usepackage{multirow,diagbox}
\usepackage[ruled,noline,noend,linesnumbered]{algorithm2e}
\usepackage[T1]{fontenc}
\usepackage[utf8]{inputenc}

\usepackage{microtype}
\usepackage{amsthm}
\usepackage{amsmath}
\newtheorem{definition}{Definition}
\usepackage{comment}
\usepackage{soul}

\usepackage{hyperref}

\usepackage[colorinlistoftodos]{todonotes}

\newcommand{\subhead}[1]{\vspace {0.03in}\noindent{\textbf{#1.}}}

\title{Does Prompt-Tuning Language Model Ensure Privacy?}

\author{Shangyu Xie\thanks{Work done during internship at Microsoft Research.}\\ \texttt{shanxie6622@gmail.com} \And Wei Dai\thanks{Work done during employment at Microsoft Research.}\\ TikTok  \\ \texttt{weidai.d@tiktok.com}\AND Esha Ghosh \and Sambuddha Roy \and Dan Schwartz \and Kim Laine \\ Microsoft Research\\ \texttt{\{Esha.Ghosh;Sambuddha.Roy;Dan.Schwartz;Kim.Laine\}@microsoft.com}\\
        }

\begin{document}
\maketitle
\begin{abstract}

Prompt-tuning has received attention as an efficient tuning method in the language domain, \emph{i.e.}, tuning a \emph{prompt} that is a few tokens long, while keeping the large language model frozen, yet achieving comparable performance with conventional fine-tuning.
Considering the emerging privacy concerns with language models, we initiate the study of privacy leakage in the setting of prompt-tuning. We first describe a real-world email service pipeline to provide customized output for various users via prompt-tuning. Then we propose a novel privacy attack framework to infer users' private information by exploiting the prompt module with user-specific signals. We conduct a comprehensive privacy evaluation on the target pipeline to demonstrate the potential leakage from prompt-tuning. The results also demonstrate the effectiveness of the proposed attack\footnote{Code available in \url{https://github.com/xiehahha/MSR_prompt_model_privacy_attack}}. 
   
\end{abstract}

\section{Introduction}
\label{sec:intro}

Large pretrained language models (LMs), \emph{e.g.}, BERT~\cite{devlin-etal-2019-bert} and GPT-2~\cite{radford2019language} have been fine-tuned for various downstream tasks to boost performance, such as dialog generation~\cite{zhang-etal-2020-dialogpt}, machine translation~\cite{fan2021beyond}, and text summarization~\cite{khandelwal2019sample}.
However, LMs-based systems have been shown to be vulnerable against various well-designed privacy attacks \cite{song2020information,hisamoto-etal-2020-membership, carlini2020extracting} and thus leak private data.
Such leakage is mainly caused by memorization in the model~\cite{236216,carlini2020extracting}.

Recently, prompt-tuning \cite{li-liang-2021-prefix,lester-etal-2021-power,liu2021pre} has achieved promising performance by only updating a tunable prompt of relatively small size (usually as a prefix to the input), while keeping the parameters of pretrained LMs frozen. 
We exploit one of the state-of-the-art prompt-tuning methods, prefix-tuning \cite{li-liang-2021-prefix}, to facilitate an simulated email service pipeline, where the various user-specific prompts will steer the LM to generate customized replies. Specifically, we integrate users' own signals in a tunable user prompt module, after which every user will have a unique prompt by prompt-tuning on the users' dataset. The results validate the effectiveness of prompt-tuning to be comparable in performance with fine-tuning (\autoref{tab:fine}).

With respect to privacy, our work aims to focus on the leakage from such prompt-tuning model. We conduct an empirical and comprehensive privacy study on a real email service pipeline with prompt-tuning. Then we design a general inference attack framework to extract private information from users. To the best of our knowledge, our work is the first study to demonstrate the privacy leakage under the prompt-tuning setting in practice.
The extensive experiments demonstrate the effectiveness of the proposed attack and potential privacy risks of the prompt-tuning model.
Above all, the proposed attack could serve as privacy evaluation to real pipeline and motivate more privacy-enhancing research in language domain.

\section{Related Work}

\begin{figure*}[!tbh]
\small
	\centering{
		\includegraphics[angle=0, width=0.9\linewidth]{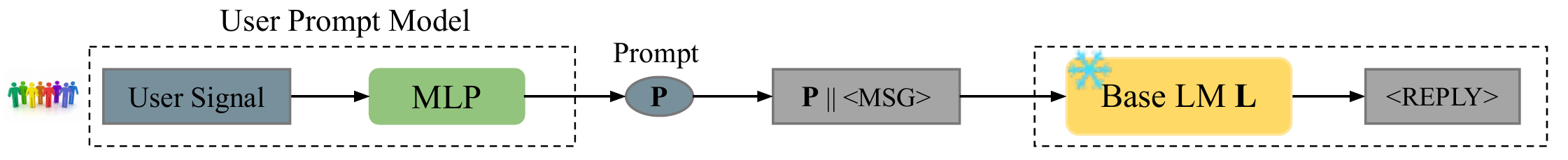}}
	\caption[Optional caption for list of figures]
	{Overview of Prompt-tuning based Email Pipeline.}
	\label{fig:framework}
\end{figure*}
\subhead{Fine-tuning}. Fine-tuning is a state-of-the-art technique for updating a pretrained large LM for an application-specific task \cite{devlin-etal-2019-bert, zhong-etal-2020-extractive,JMLR:v21:20-074}.
For example, there have been several works for text summarization via fine-tuning on different models, such as masked language model BERT \cite{devlin-etal-2019-bert}, encoder-decoder BART \cite{lewis-etal-2020-bart}, and decoder-only GPT model~\cite{radford2019language, khandelwal2019sample}.
Fine-tuning technique is also the main paradigm on language generation, \emph{e.g.}, dialog generation \cite{zhang-etal-2020-dialogpt},and machine translation~\cite{fan2021beyond}.

\subhead{Prompting and Prompt-tuning}
Due to the high computational overhead in updating all model parameters, recently lightweight tuning methods, such as prompt-tuning \cite{li-liang-2021-prefix,lester-etal-2021-power,zhong-etal-2021-factual} have received a lot of attention.
Prompt-tuning is motivated by prompting, where additional information is added as a condition to steer the language model to generate specific kind of output.
For instance, GPT-3 adopts a manually designed prompt for the generation of various tasks \cite{NEURIPS2020_1457c0d6} as prompt engineering.
Other prompt-engineering approaches include manual or non-differentiable search methods~\cite{jiang2020, shin-etal-2020-autoprompt}. 
Prompt-tuning steps further convert the design of a prompt to the optimization in a continuous space, \emph{e.g.}, a continuous prefix embedding \cite{li-liang-2021-prefix}, which can be more expressive and achieve comparable performance with fine-tuning.

\subhead{Privacy Attacks in Language} Membership inference attacks (MIAs) \cite{shokri2017membership,hisamoto-etal-2020-membership} aim to determine whether a data sample was used or not in the training set of the target model (as membership), which are identified as the state-of-the-arts privacy attacks to evaluate the privacy leakage in ML domain due to its simplicity. There are several MIA works concerned with NLP systems, such as translation \cite{hisamoto-etal-2020-membership} and sentence classification~\cite{mahloujifar2021membership,mireshghallah2022quantifying}. Besides, data reconstruction attacks \cite{song2020information,carlini2020extracting,zhu_deepleak} aim to recover or to extract private information from the ML pipeline with the auxiliary knowledge of the model. For example, Carlini \emph{et al.}~\cite{carlini2020extracting} first demonstrated that the attacker can extract training data from a language model by directly querying the model due to the memorization~\cite{236216}. Different from previous attacks on fine-tuning, we focus on the prompt-tuning model for a real-world email service pipeline and design privacy attacks (\autoref{sec:threat}).

\section{Target Email Service Pipeline} \label{sec:bg}

In this section, we demonstrate our prompt-tuning email service pipeline.

\autoref{fig:framework} depicts the framework of our proposed prompt-tuning pipeline. 
We focus on a real-world email service pipeline which provides customized replies automatically, \emph{i.e.}, the pipeline will auto-complete diverse replies (denoted as ``\texttt{<REPLY>}'') for users given the same messages (denoted as ``\texttt{<MSG>}'') on user's individual information accordingly.
There are two main modules in the pipeline: 

\vspace{0.05in}
\noindent\textbf{1) User Prompt Model}: aims to control the user-specific prompt (denoted as $P$) with the input of user information (\emph{e.g.}, writing raw data).
We construct a unified MLP model (to avoid unstable optimization \cite{li-liang-2021-prefix}) as User Prompt Model. Such unified model is different from the original prefix-tuning method \cite{li-liang-2021-prefix} to train a separate MLP model for every user to obtain customized output, which can bring costly computational overhead in large-scale deployments. Our unified-model adaptation could match better with a real-world system in both computational overhead and scalability. Specifically, we define an N-gram feature vector as the input signal (denote as $u$) for every user, which represents the frequency of a fixed N-grams dictionary from the original user's email corpus.\footnote{The reason we select such an N-gram feature vector as user's signal is that it could reflect the difference/similarity of an individual's profile by the distribution of N-grams approximately~\cite{damashek1995gauging, wang2007topical}.} Such fixed N-grams dictionary of size $k$ consists of top-$k$ frequent N-grams of the entire users' email corpus. We will get prompt: $P=\texttt{MLP}_{\theta}(u)$ ($\theta$ is the MLP parameters).

\vspace{0.05in}
\noindent\textbf{2) Base LMs}: the prompt $P$ will be prepended to the <MSG> to steer LMs (denoted as $L$) to generate customized replies <REPLY> using prompt-tuning:
\begin{equation}
\texttt{<REPLY>}=\mathcal{L}(P ~||~ \texttt{<MSG>})
\end{equation}

\vspace{0.05in}
\noindent\textbf{Privacy Implications}.
In practice, we will leverage distributed learning techniques, such as Federated Learning \cite{mcmahan2017communication}, to train the pipeline.
Specifically, each user's personal emails will be locally stored and computed during the training process, only sending gradients to update the MLP model, so that the only information disclosed to the our server are the N-gram feature vectors (for generating replies).
\cite{li-liang-2021-prefix} indicates that a prompt-tuning based pipeline could obtain intrinsic privacy protection for individual users. In this work, we show that prompt-tuning has non-trivial privacy implications for the user.

\section{Privacy Attack Framework}
In this section, we first introduce the threat model, including the attacker's capabilities and then detailed our attack methodology.

\subsection{Threat Model}\label{sec:threat}

We first formulate our privacy threat model, including the attack scenarios, attacker's goal and capability. We define three entities: 

\noindent (1) \textbf{Service Provider}: The service provider will build a prompt-tuning based pipeline to provide better email service for its users, the back-end model being trained on private email data collected from users. The service provides a public API for the auto-complete function for each user.

\noindent (2) \textbf{Target User}: Users are usually from a specific organization, such as a school or a hospital. The users' email inboxes can contain highly confidential data, such as business-related information. These messages are utilized to train the model.
    
\noindent (3) \textbf{Attacker}: The attacker may have access to the email service as a normal user and can query the API to get the output of the pipeline, which thus infer private information (attacker's goal). Note that there may exist different cases based on the role of attackers, thus we define two types of attacks: 1) \emph{third-person} attack; 2) \emph{first-person} attack. For example, the attacker may be undercover in a specific organization to infer sensitive information about the organization or other users (as a third-person attack). 

In addition, an attacker may also be a ``good'' stakeholder. For instance, the organization may want to check whether their private data is abused or collected by the service provider, in conflict with business contracts, licenses, or data regulation. In this case, such a stakeholder may leverage our proposed attack as a first-person attack.

\subhead{Attacker's Capability and Knowledge} Our attack works in a black-box setting, \emph{i.e.}, the attacker does not access to the models and their parameters. Fake as good user, the attackers can get as many outputs by querying the targeted pipeline as they want. Specifically, the attacker can set up another account to send emails to itself and collect the outputs from the pipeline.

\begin{figure}[!tbh]
	\centering{
		\includegraphics[angle=0, width=0.95\linewidth]{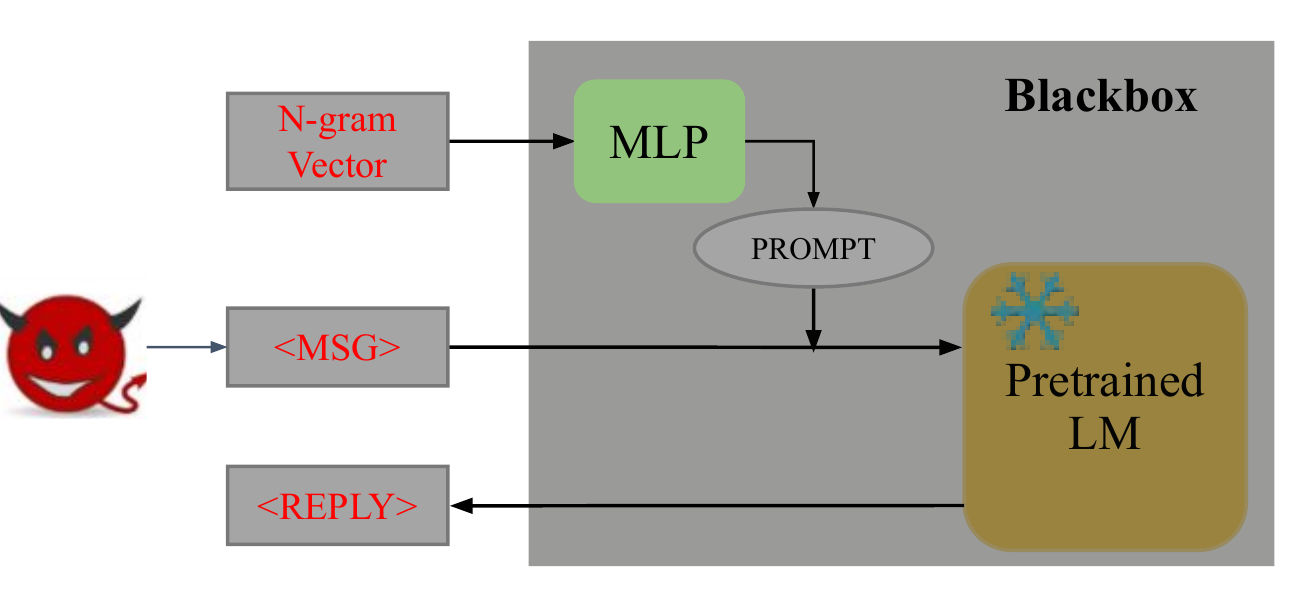}}
	\caption[Optional caption for list of figures]
	{Proposed Attack. Attacker will compute N-gram vector of target user and query the pipeline with multiple messages (<MSG>) to get customized replies (<REPLY>), thus inferring private data.} 
	\label{fig:att_f}
\end{figure}

\subsection{Attack Methodology}\label{sec:attck_method}

The main idea of the proposed attack would be targeting \emph{user prompt} model in the pipeline, which can produce a user-specific prompt with user signals. That is, if the attacker can possibly get the target user's feature vector, or utilize some auxiliary knowledge to get partial information of such feature vector, it would be more likely that the attacker could extract the private information from a target user (overview in \autoref{fig:att_f}).

For the first-person attack, the attacker would directly know the target user's signal. For the third-person case, the attacker cannot get the original target user's signal but could yield partial knowledge. As a close contact in the organization, the third-person attacker may communicate with the target user to collect a (partial) set of target user's replies and try to impersonate the target user to fool the pipeline to return customized output.

\subhead{Privacy Notion}
To better track the private information in the user's email corpus, we would inject random tokens (usually infrequently used words referring as ``carnaries'') to the training set as private information and then check the existence of such private words.
This is a commonly used method for modeling private information~\cite{236216, song2019auditing,leakgeccs21, elmahdy-etal-2022-privacy}.
This also matches with our defined threat model, \emph{i.e.}, first-person attack can be utilized to detect the unauthorized usage/collection of data~\cite{song2019auditing}. 

\begin{definition}[Leakage] If the target user's canaries are included in the decoded query output of the pipeline, the output will be viewed as leakage.
\end{definition}

Besides, we define the following notion to help the quantitative measurement of leakage. Given a collection of one target user's training data, we inject private tokens into the replies at random locations with the two privacy parameters: 1) $\alpha_{{reply}}$, the percentage of inserted replies with private tokens out of the total user's replies; 2) $\alpha_{{word}}$, the percentage of inserted private tokens out of one single reply. Such private token-augmented user data will be used as a part of the training dataset for our pipeline. 
With aspect of third-person attack, we also define a partial knowledge parameter $\beta$, as the percentage of the data collected by third-person attacker in the target user's email corpus, which will be used to recover the target user's N-gram feature vector (if $\beta=1$, it is first-person attack).

\subhead{Third-Person Attack}
In most cases, the attacker is not the target user.
This indicates that the attacker will try to recover the user N-gram feature vector and infer the private tokens as following steps:

\subhead{Step (1)} The attacker first will get as many as target user's raw replies/emails, which is utilized to compute the target user's feature vector.
    The more data the attacker collects (higher $\beta$), the more precise the derived feature vector will be.

\subhead{Step (2)} The attacker will designate another email account to fake the target user, \emph{e.g.}, the attacker can compose lots of email message-reply pairs to be collected by the pipeline.
    The pipeline may produce customized output similar to the target user, since the designated account displays a similar user feature vector to the target user.
    In fact, our experimental results shows that such partial knowledge can serve as a strong factor to infer the private tokens as third-person (\autoref{exp:attack}.)

\subhead{First-Person Attack}
As the attacker is actually the target user (first-person), the attacker can directly query the pipeline to check whether the output includes the private tokens.
The privacy parameters $\alpha_{{reply}}$ and $\alpha_{{word}}$ can be tuned to present a more quantitative result.

In both attacks, the attacker will query the pipeline with randomly selected messages to generate replies and check if there is leakage. We compute the leakage rate that refers to the percentage of leakage out of total queries. Note that the choice of decoding method can contribute the leakage. We utilize greedy decoding by default (select the top-1 token). Optionally, we can also choose beam search if the output includes logits, which can potentially cause more leakage, as shown in ablation study (Section \ref{exp:attack}).

\section{Experimental Evaluation}
\begin{figure*}[!tbh]
	\centering
	\subfigure[Private Word]{
		\includegraphics[angle=0, width=0.249\linewidth]{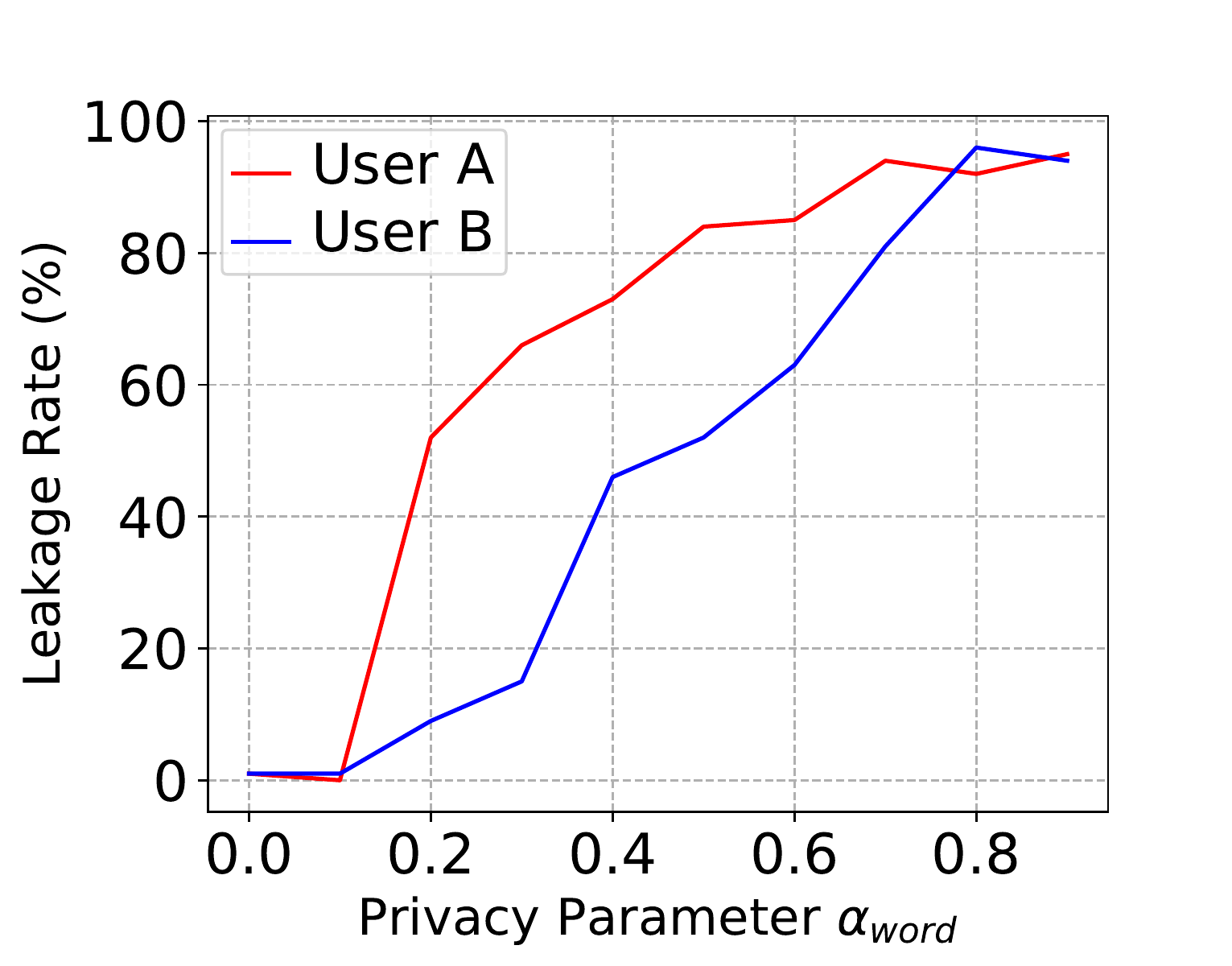}
		\label{fig:l-1} }\hspace{-0.2in}
		\subfigure[Private Reply]{
		\includegraphics[angle=0, width=0.249\linewidth]{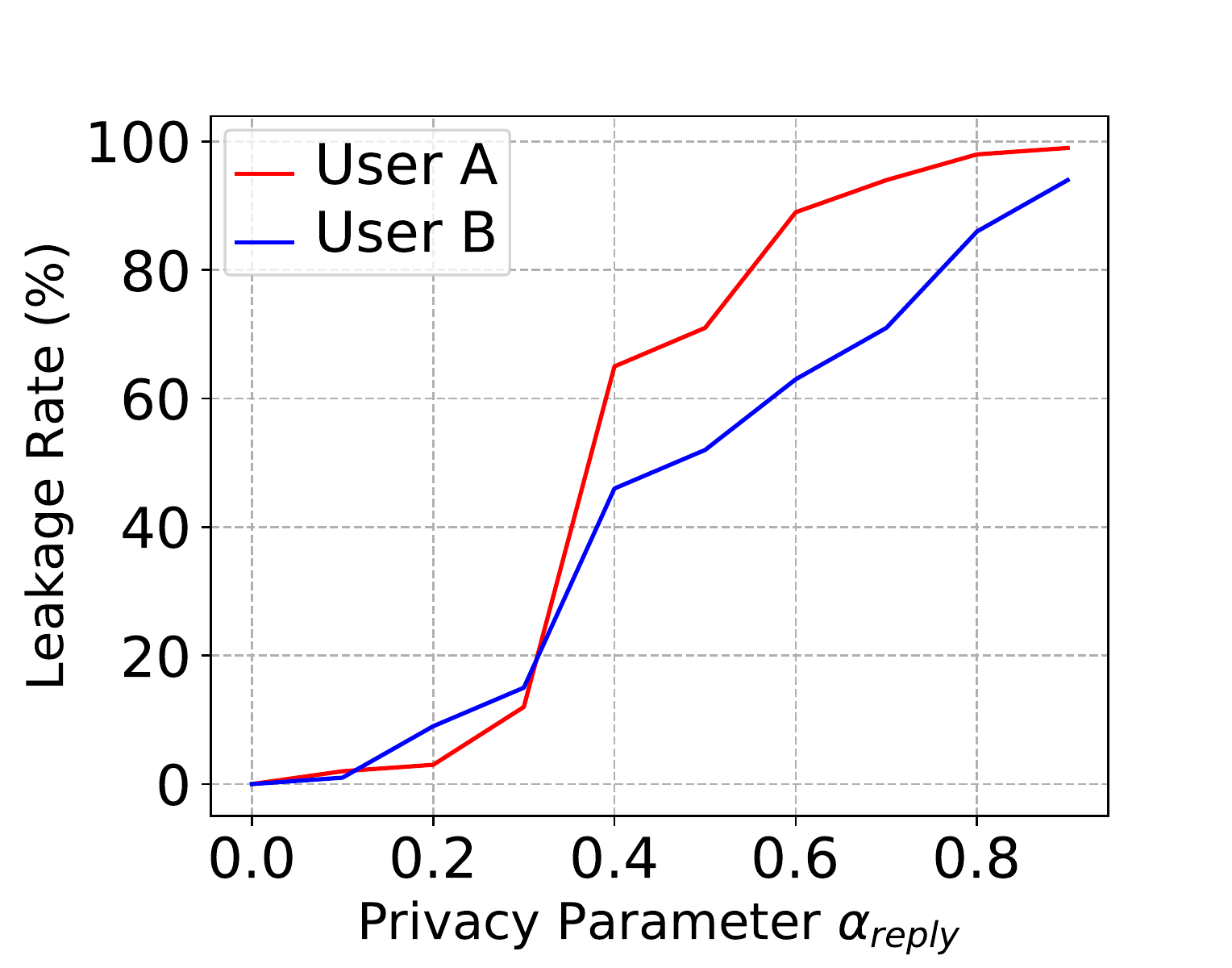}
		\label{fig:l-2} }\hspace{-0.2in}
		\subfigure[Partial Parameter]{
		\includegraphics[angle=0, width=0.249\linewidth]{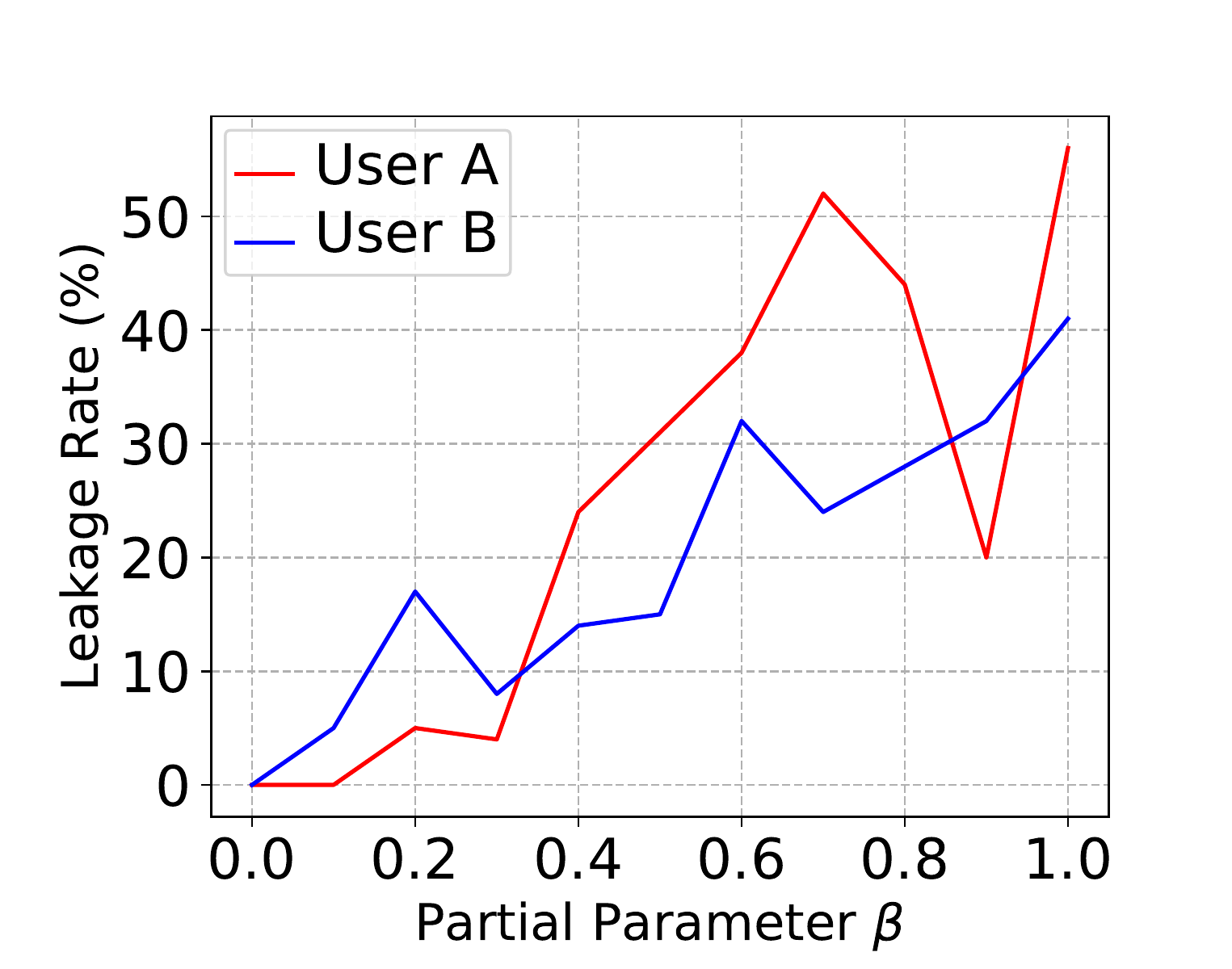}
		\label{fig:l-3}}\hspace{-0.2in}
		\subfigure[Training Epochs]{
		\includegraphics[angle=0, width=0.249\linewidth]{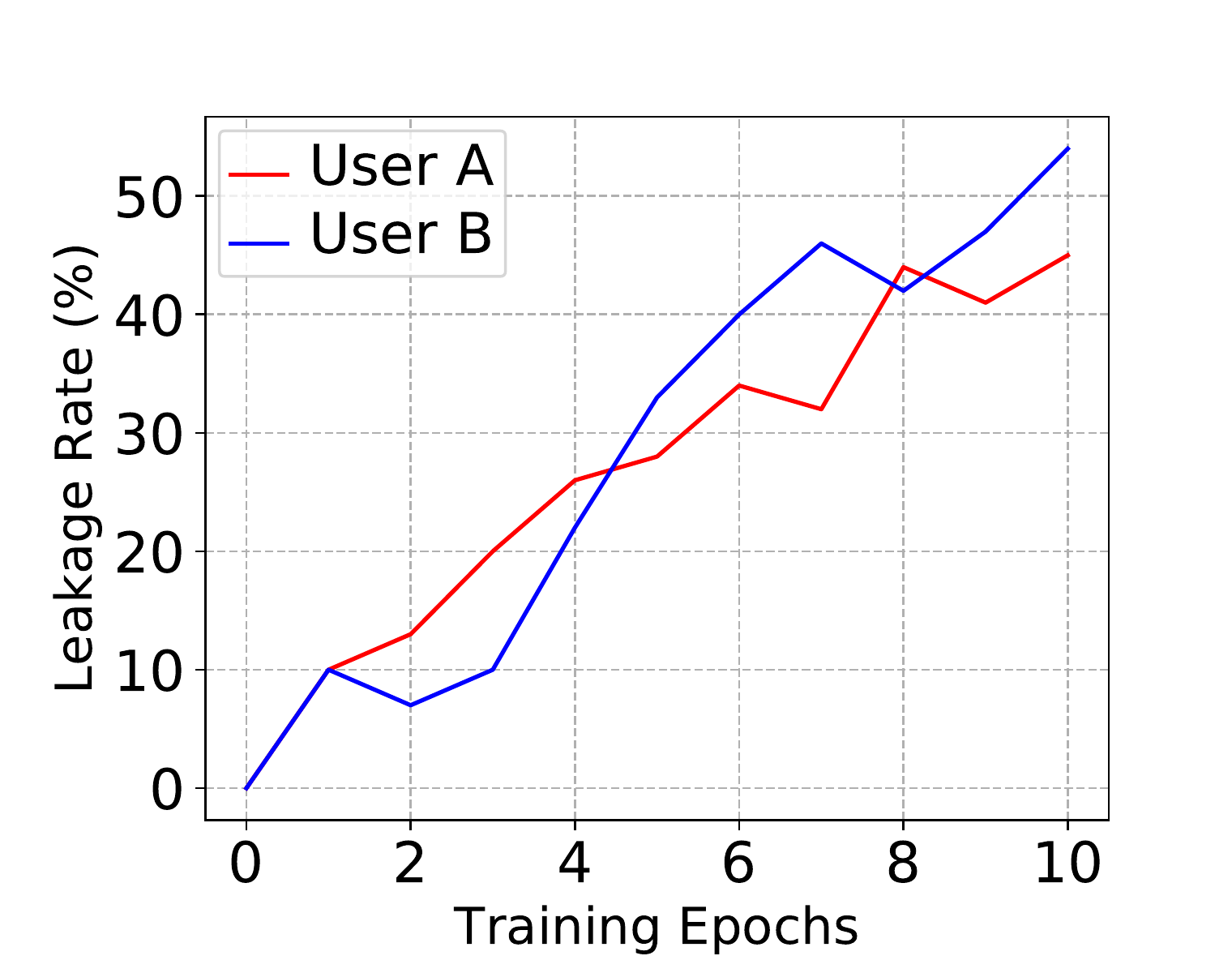}
		\label{fig:l-5} }\hspace{-0.2in}
		\vspace{-0.1in}
	\caption[Optional caption for list of figures]
	{Leakage vs. Different Privacy parameters (abc), Number of Training Epochs (d).} \label{fig:leak}
\end{figure*}

\subsection{Experimental Setup}

\subhead{Dataset and Model}
We first evaluate the prefix-tuning method on a real but defunct company email dataset, the Avocado Research Email Collection \cite{avocado} which includes emails from about $280$ users' email accounts.
We extract around 300K message-reply pairs from the raw dataset and divide them into training/test data with a $80:20$ split.
We adopt the OpenAI GPT-2 model \footnote{\url{https://huggingface.co/docs/transformers/model_doc/gpt2}} from Huggingface with the default BPE tokenizer. 
Note that we removed the signature of the emails to prevent the side-information leakage, \emph{e.g.}, name and job title.
We will generate the data by appending the reply (target) to the input message (source) separated with a <SEP> token. Then we only focus on the training loss of target sequences for either fine-tuning or prompt-tuning (experimental setup and results are in appendix \ref{exp:prefix}). The results have shown that such prompt-tuning can achieve comparable performance with fine-tuning. We will focus on the privacy evaluation as follows.

\subhead{Private Token}
We utilize a Diceware word list\footnote{\url{https://www.eff.org/files/2016/07/18/eff_large_wordlist.txt}} published by EFF, which contains lots of rare words to aim to improve the security of passwords.
We select a private token from this list for the convenience of our privacy check.
In this case, we choose to use the word ``appendage'' as the private token to be inserted into the target user's replies.

\subsection{Performance of Prompt-tuning}\label{exp:prefix}

We fix the length of incoming messages (source) to be $150$ tokens and replies (target) to be $50$ tokens. For fine-tuning, we use AdamW optimizer with weight decay $0.01$ and learning rate $5e^{-5}$, with $500$ warm-up steps.
The number of epochs is~$5$.
For prompt-tuning, we first extract a top-$10000$ $N$-gram count vectors ($N\leq3$) from each individual user's emails as user-domain signal.
Then we use a MLP model to project the $10000$-dimensional vector to a length $K=20$ user prompt vector (dimension $20 \times 768$ to match with GPT-2 activations), which thus works as a prefix to steer the language model to generate reply sequences.

We report the perplexity and accuracy (the percentage of correctly generated tokens out of the original replies) for both methods.
\autoref{tab:fine} demonstrates the results.
Note that this shows comparable performance for fine-tuning and prompt-tuning.

\begin{table}
\small
\centering
\begin{tabular}{c|cc}
\hline
Method &Fine-tuning & Prompt-tuning \\ \hline Loss/Perplexity & $31.75$ & $27.68$\\ \hline

Accuracy & $31.9\%$ & $32.3\%$ \\ 
\hline
\end{tabular}
\caption{\label{tab:fine} Performance of the pipeline with Fine-tuning and Prompt-tuning. The initial Huggingface as baseline has perplexity $112.5$ and accuracy $12.4\%$.}
\end{table}

\subsection{Attack Evaluation}
\label{exp:attack}

We evaluate the quantitative leakage with the privacy parameters defined for first/third-person attacks: 1) privacy parameters $\alpha_{word}, \alpha_{reply} \in [0,1)$; 2) partial knowledge parameter $\beta \in [0, 1]$, respectively. Note for each single evaluation, the attacker will query to collect 500 generated replies and report the average value of the 5 repetitions.

We first evaluate the leakage rate of varied privacy parameters (directly referring to the percentage of private tokens) from the user's original data. We fix $\beta=1$ as the first-person attack, where the attacker already knows the N-gram vector. We set $\alpha_{word}=0.2, \alpha_{reply}=0.3$ for every subgroup experiments unless specified, which makes the percentage of private tokens very small ($\sim 0.1\%$) out of the whole training set. We have tested the leakage of 50 different users as target user independently (results in appendix \ref{sec:appendix-leakge}). and selected two random users, denoted ``User A'' and ``User B'' for better demonstration. Figure \autoref{fig:l-1} (fixing $\alpha_{reply}=0.3$) demonstrates the leakage with varying $\alpha_{word}$. We can observe that the leakage increases as privacy parameter increases. Figure \autoref{fig:l-2} (fixing $\alpha_{word}=0.2$) shows similar trend. These are expected since we have manually increased the number of private tokens in the dataset (which are inclined to be memorized by the pipeline).

Next, we test the third-person inference attack, which is more devastating as the attacker may be able to infer previously unknown private information of target users. We report the leakage with varying partial parameter $\beta \in [0, 1]$. As shown in Figure \autoref{fig:l-3}, the leakage increases as the partial parameter $\beta$ increases. This is reasonable that the attacker can get more accurate target user's N-gram vector with more collected email corpus. Such results show that the pipeline could still leak the private information even if the attacker does not directly obtain the original N-gram feature vector. One possible solution is that the service provider can build up an effective spam filter, since the attacker would send emails to try to collect user's replies to obtain the target user's N-gram vector. However, the attacker could be the close contact of the target user (\emph{e.g.}, belonging to the same organization), which could bypass such spam filtering.

\subsection{Ablation Study} Specifically, we will vary the studied parameter (\emph{e.g.}, training epochs) independently, while fixing other privacy parameters, and finally report the leakage rate.
We also conduct ablation studies for our proposed attack in aspect of training epochs and decoding strategy. First, we compute the leakage rate with various numbers of training epochs $[0, 10]$ (shown in Figure \autoref{fig:l-5}).
We can observe that the leakage also gets higher as training epochs increases.
This is due to the model overfitting on the training set (and thus memorizing private information). Second, we evaluate the impact of decoding strategies (greedy with top-1 output and beam Search with logits), corresponding to the attacker's knowledge.
That is, beam search will utilize full logits to generate the final replies, while the greedy strategy only chooses the top-1 output for decoding; we set the beam size to be~4.
\autoref{tab:epo} demonstrates the final leakage with the two strategies.
As expected, we find that beam search causes significantly higher leakage than greedy decoding.

\begin{table}
\small
\centering
\begin{tabular}{c|cc}
\hline
 &Greedy & Beam Search\\ \hline 
User A & $0.41$ & $0.68$ ($\uparrow .27$) \\ \hline
User B & $0.35$ & $0.53$ ($\uparrow .18$) \\ 
\hline

\end{tabular}
\caption{\label{tab:epo} Leakage vs. Decoding Method.}
\end{table}

\section{Conclusion}
In this paper, we design a privacy attack targeting prompt-tuning, which can infer private information from a real email service pipeline. With carefully designed attacks, we conduct a comprehensive privacy evaluation to the pipeline to show the potential leakage. Experimental results demonstrate the effectiveness of the proposed attack and also comply with the previous theory: memorization also exists in prompt-tuning models.

\section{Limitations}

The proposed attack is limited in practice due to a ``strong'' attack setting (defined in threat model \autoref{sec:threat}). Specifically, the attacker has good privileges and enough knowledge to infer the target user's signal. For instance, the attacker can fake as a normal user to receive the emails and even extract the private messages from the targeted users. Besides, the attackers could also query the pipeline as many times as they can. We argue that this is reasonable setting considering the feasibility of being an internal user belong with an organization. In addition, for a privacy evaluation study to demonstrate the leakage, we should always consider a stronger attack, which is similar to the defense evaluation (against stronger instead of weaker attack). Such setting has been utilized in previous machine learning security works \cite{tramer2020adaptive,athalye2018robustness}. Another limitation is the choice of private tokens. For better quantitative evaluation, we select the rare word, e.g., ``appendage'', which can increase the leakage to some extent since the training loss will focus on such rare sequences (more memorization). We omit the effect of insert position as random insertion. We will conduct more experiments in aspect of the private tokens as future work.

In aspect of defense against the proposed attack, one potential method is to apply differential privacy (DP) \cite{dwork2006calibrating, yu2021differentially} during the training of pipeline, which has been shown to be effective in protecting data privacy.
However, there exists a utility-privacy trade-off, where the DP noise prevents the model from capturing the general data distribution and thus degrades utility. Another feasible method would be auditing the training dataset for the pipeline, \emph{e.g.}, we could manually filter out personal information from the dataset, which may require lots of engineering labor and work, as the dataset for training such large models is usually very large. We will look into these directions in the future.

\section{Ethical Statements}

In this work, we propose a privacy attack to empirically evaluate a real-world pipeline with the prompt-tuning to reveal potential privacy leakage.
Such privacy-related work should be considered to be more ethical than harmful. Our attack can serve as an internal red teaming technique to demonstrate the privacy risks to stakeholders.
Finally, we believe our attack can motivate new types of privacy-enhancing works in the language domain.

% \section*{Acknowledgements}

% This document has been adapted
% by Steven Bethard, Ryan Cotterell and Rui Yan
% from the instructions for earlier ACL and NAACL proceedings, including those for 
% ACL 2019 by Douwe Kiela and Ivan Vuli\'{c},
% NAACL 2019 by Stephanie Lukin and Alla Roskovskaya, 
% ACL 2018 by Shay Cohen, Kevin Gimpel, and Wei Lu, 
% NAACL 2018 by Margaret Mitchell and Stephanie Lukin,
% Bib\TeX{} suggestions for (NA)ACL 2017/2018 from Jason Eisner,
% ACL 2017 by Dan Gildea and Min-Yen Kan, 
% NAACL 2017 by Margaret Mitchell, 
% ACL 2012 by Maggie Li and Michael White, 
% ACL 2010 by Jing-Shin Chang and Philipp Koehn, 
% ACL 2008 by Johanna D. Moore, Simone Teufel, James Allan, and Sadaoki Furui, 
% ACL 2005 by Hwee Tou Ng and Kemal Oflazer, 
% ACL 2002 by Eugene Charniak and Dekang Lin, 
% and earlier ACL and EACL formats written by several people, including
% John Chen, Henry S. Thompson and Donald Walker.
% Additional elements were taken from the formatting instructions of the \emph{International Joint Conference on Artificial Intelligence} and the \emph{Conference on Computer Vision and Pattern Recognition}.


\bibliography{anthology,custom}
\bibliographystyle{acl_natbib}

\appendix

\section{Leakage of Varied Users}
\label{sec:appendix-leakge}
\begin{figure}[!tbh]
	\centering{		\includegraphics[angle=0, width=0.8\linewidth]{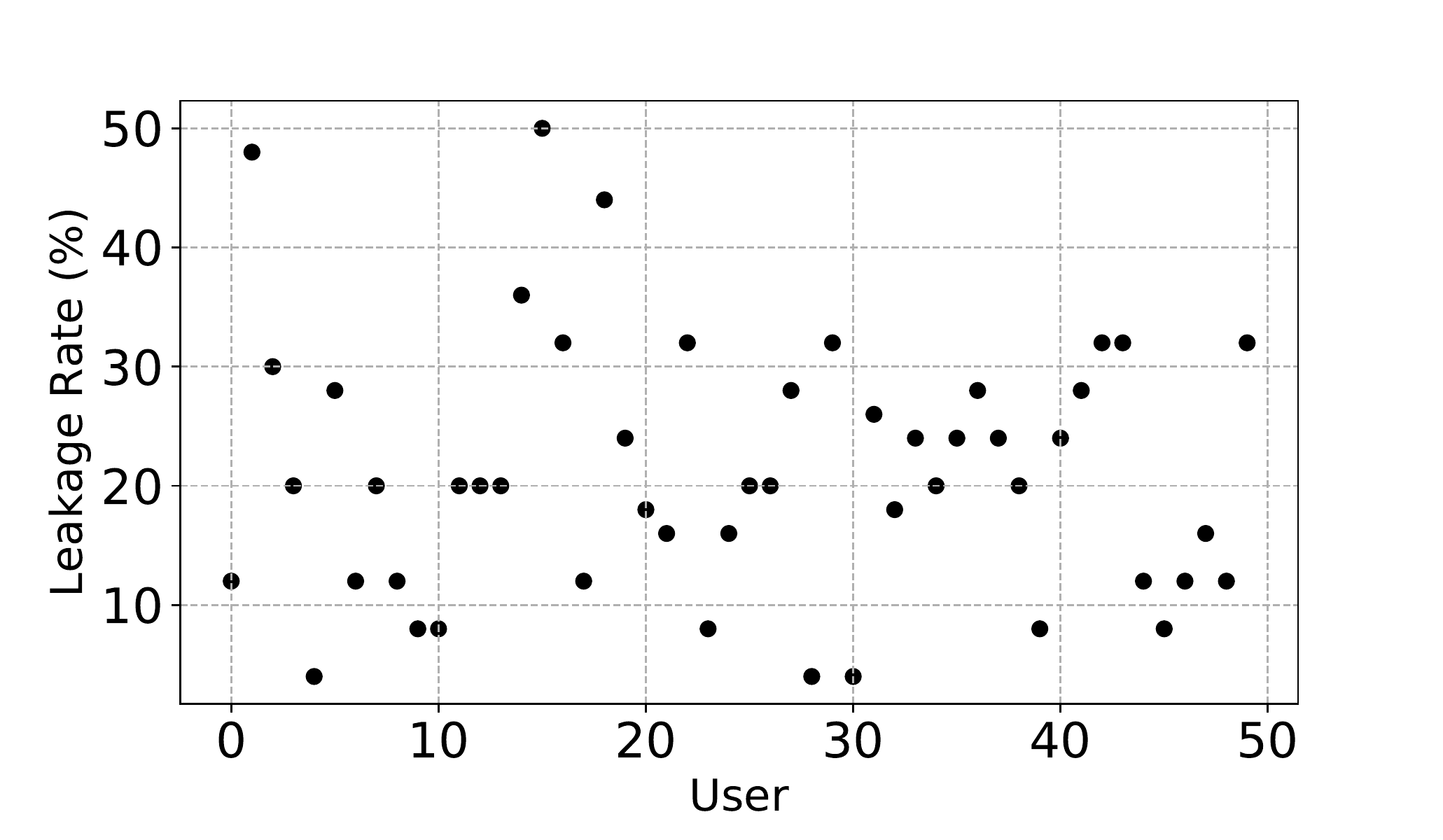}}
		\vspace{-0.1in}
	\caption[Optional caption for list of figures]
	{Leakage of Varied Users.} 
	\label{fig:user}
\end{figure}

\autoref{fig:user} demonstrates the leakage of varied users against our proposed privacy attack, which also validates the effectiveness of attack. We observe that different users could have different leakage, which is reasonable since some users may obtain larger set of messages/replies, \emph{e.g.,} stakeholders or managers. Note that our attack can be readily extended to multiple users (\emph{i.e.}, set up multiple designated account to obtain the target users' N-gram feature vector separately. 

\end{document}